\documentclass[twocolumn,showpacs,preprintnumbers,amsmath,amssymb,pre]{revtex4}

\usepackage{graphicx}% Include figure files

\begin{document}

\preprint{}

\title{Random walk and trapping processes on scale-free networks} 

\author{Lazaros K. Gallos}
\affiliation{Department of Physics, University of Thessaloniki, 54124 Thessaloniki, Greece}

\date{\today}

\begin{abstract}
In this work we investigate the dynamics of random walk processes on scale-free networks
in a short to moderate time scale.
We perform extensive simulations for the calculation of the mean squared displacement,
the network coverage and the survival probability on a network with a concentration $c$
of static traps. We show that the random walkers remain close to their origin, but cover
a large part of the network at the same time. This behavior is markedly different than
usual random walk processes in the literature. For the trapping problem we numerically
compute $\Phi(n,c)$, the survival probability of mobile species at time $n$, as a function of the concentration
of trap nodes, $c$. Comparison of our results to the Rosenstock approximation indicate that this is
an adequate description for networks with $2<\gamma<3$ and yield an exponential decay. For $\gamma>3$
the behavior is more complicated and one needs to employ a truncated cumulant expansion.

\end{abstract}


\maketitle

\section{Introduction}
\label{sect:intro}

Scale-free networks have been widely studied during the recent years, mainly because of their
connection to a plethora of real-world structures \cite{Albert,Dorogovtsev}. These networks are made
by nodes connected to each other via links, which may be directed or undirected (in the present
work we only deal with undirected links). Studies of their structure show that most of them possess the small-world property,
i.e. the mean path length is extremely small and every node can be reached by following a
path consisting from a very small number of nodes, as compared to the case of lattice systems.
A special feature, though, that distinguishes this class of networks is the fact that the probability
distribution for a node to have $k$ links to other nodes obeys a power law:
\begin{equation} 
P(k) \sim k^{-\gamma} \, ,
\end{equation}
where $\gamma$ is a parameter that measures how densely connected a network can be. There is a wide range
of real-life networks \cite{Albert,Dorogovtsev} that have been shown to follow this power-law form in their connectivity, including
networks in nature, such as the cell, metabolic networks and the food web, artificial networks such as the Internet, the WWW and power grids,
or even social networks, such as sexual partnership networks.

The scale-free networks, termed after the absence of a characteristic typical node connectivity, exhibit
many unusual properties as compared to simple lattice models, random graphs, or even small-world (Watts-Strogatz \cite{WS}) networks.
This scale-free character results in the existence of a small number of nodes which are connected to a large
number of other nodes. These super-connected nodes (termed `hubs') have been shown to have a central place
in the interpretation of many of the network properties.
A lot of work has been devoted in the literature to the study of static properties of the networks,
while interest is growing for dynamical properties on these networks.
Recently \cite{GA04}, we presented results for the absence of kinetic effects in reaction-diffusion
processes taking place on scale-free networks.
In this work, we study a number of random walk properties, including mean-squared displacement, network coverage, and
trapping processes on scale-free networks of varying connectivities. Trapping has been
considered in the past as a model for energy transfer, but also in a more general frame in the context of networks,
as a model for the probability of reaching targets located on the network in a
given concentration via random moves \cite{ALPH}.
It is of interest, thus, to study the mechanism, the effects of connectivity, concentration, size, etc on such structures
that exhibit these unique properties. Our results refer to small to moderate time regimes, where we are still far from the
asymptotic limit. This limit has been known to be very hard to reach in regular lattices, too, and cannot be predicted by
direct simulation techniques.

\section{Random walks and the trapping problem}

One of the most basic quantities in the random walk theory is the mean-squared displacement $\langle R^2(n) \rangle$ of a particle diffusing
in a given space, which is a measure of the distance $R$ covered by a typical random walker after performing $n$ steps. In most cases,
this quantity is described by an expression of the form:
\begin{equation}
\label{EQmsd}
\langle R^2(n) \rangle \sim n^a \,.
\end{equation}
The value of the parameter $a$ classifies the type of diffusion into normal linear diffusion ($a=1$), sub-diffusion ($a<1$),
or super-linear diffusion ($a>1$). Of course, when we consider distinct time steps and nearest neighbor lattice hops the maximum value of
$a$ can be 2, i.e. a completely biased walk where the random walker continuously moves away from the origin.
Recently \cite{Almaas}, the mean squared displacement was studied in small world networks, where it was shown that
diffusion is linear and results were found to collapse under a proposed scaling.

The behavior of a random walk is also characterized by the coverage of the space, as expressed by the average number of
distinct sites visited $\langle S_n \rangle$ after $n$ steps. In regular Euclidean lattices this quantity follows a power law with the
number of steps, except in the case of two dimensions, where logarithmic corrections appear in the denominator
($\langle S_n \rangle \sim n /\ln{n}$). In one dimension $\langle S_n \rangle \sim \sqrt{n}$, and in
dimensions higher than two $\langle S_n \rangle \sim n$, and the number of sites visited grows linearly with the number of steps $n$, since the
random walker practically visits at each step a new site. In infinite dimensions, of course, the number $\langle S_n \rangle$ of visited sites
is equal to $n$, since there is no revisitation of sites during the walk, and the walker covers the largest possible area.
In small-world networks a scaling ansatz was proposed \cite{Almaas}, which was verified by simulations, and $\langle S_n \rangle$
shows a transition from a slope 0.5 (one-dimensional behavior) to a slope 1 ($d>2$ behavior).

An important process related to random walk theory is trapping \cite{WeissBook,WeissReview}.
Trapping reactions have been widely studied in the frame of physical chemistry, as part of the general
reaction-diffusion scheme. The general idea includes two different species A and B, which diffuse freely in a given space and upon proximity
they react according to A+B$\to$B. Many different variations describe a plethora of physical phenomena.
In this paper we deal with the special case of the trapping problem where B particles are immobile.
The simplest mean-field analytical treatment predicts a simple exponential decay in the density of A's,
while the earlier contributions to the subject go back to Smoluchowski \cite{Smoluchowski},
who was the first to attempt to relate the macroscopic behavior with the microscopic picture
by taking into account local density fluctuations. However, over the years a lot of work \cite{WeissBook,WeissReview} has been
devoted to the trapping problem which, even in its simplest form, was shown to yield a rich diversity of
results, with varying behavior over different geometries, dimensionalities and time regimes.

The main property monitored during such a process is the survival probability $\Phi(n,c)$, which denotes
the probability that a particle A survives after performing $n$ steps in a space which includes traps B with a concentration $c$.
It is well-known that $\Phi$ behaves differently in different dimensions, as well as in different time-regimes.
The problem was studied in regular lattices and in fractal spaces\cite{WeissBook,WeissReview}, and, recently, in small-world networks
by Blumen and Jasch \cite{Jasch2001,Blumen2002,Jasch2002}.

The simplest treatment of the trapping problem on a lattice assumes that when a random walker has performed $n$ steps
and has visited $S_n$ different lattice sites at least once, the probability that it has not yet been trapped is equal to $(1-c)^{S_n}$,
where $c$ is the trap concentration. When this quantity is averaged over all different possible walks
and trap configurations the resulting survival probability will be equal to
\begin{equation}
\label{exact}
\Phi(n,c) = \left\langle (1-c)^{S_n} \right\rangle = \left\langle e^{-\lambda S_n} \right\rangle \, ,
\end{equation}
where $\lambda=-\ln(1-c)$. A simplification of this equation was first proposed by Rosenstock \cite{Rosenstock} and
simply substitutes the above expression with the typical value of the distribution, i.e.
\begin{equation}
\label{Rosen}
\Phi(n,c) = e^{-\lambda \left\langle S_n \right\rangle} \, .
\end{equation}
This approximation has the advantage that the mean value of the number of sites visited $\left\langle S_n \right\rangle$
is well known \cite{Montroll} for practically all dimensionalities (including e.g. fractal ones).
Notice that the Rosenstock approximation does not necessarily imply a simple exponential decay, except in the
case where $\left\langle S_n \right\rangle \sim n$.
The formula predicts simple exponential decay of the survival probability with the number of steps $n$ only for $d\geq 3$,
and exponential dependence on $\sqrt{n}$ in $d=1$. In 2 dimensions the predicted behavior is rather complex, with
logarithmic corrections in the exponent.
The applicability of Eq.~(\ref{Rosen}) is limited to short-times and/or not too large trap concentrations.
When the survival probability becomes low enough, this expression deviates significantly from the correct behavior.

A significant improvement was possible by the use of averaged quantities, known as cumulants, where the averaged
quantity of Eq.~(\ref{exact}) can be written as a function of
the cumulant generating function \cite{BlumenCumulant}:
\begin{equation}
\label{cum_gf}
K_J(\lambda,n) = \sum_{j=1}^{J} (-1)^j \frac{\lambda^j}{j!} k_j(n) \;\;\; ,
\end{equation}
where $k_j(n)$ are the cumulants, which are associated to the moments of $S_n$, e.g.
$k_1(n)=\langle S_n \rangle$, $k_2(n)=\langle S_n^2 \rangle - \langle S_n \rangle^2$, etc.
The expression (\ref{exact}) for the survival probability then simply becomes
\begin{equation}
\label{EQ_cum_surv}
\Phi_J(n,c) = \exp (K_J(\lambda,n)) \;\;\; .
\end{equation}
Improved accuracy can be obtained upon increasing the truncation order $J$. In theory,
the knowledge of all the moments ($J\to \infty$) for the $S_n$ distribution is required for the use of (\ref{cum_gf}).
These moments are known analytically only in one-dimensional lattices, while for $d>1$ usually the first 2-4 moments are used.

A detailed analytical treatment of the problem was performed by Donsker and Varadhan \cite{DV}, who were able
to predict the asymptotic behavior of the survival probability as
\begin{equation}
\label{dv}
\lim_{n\to\infty}\Phi(n,c) = \exp(-K_d \lambda^{\frac{2}{2+d}} n^{\frac{d}{d+2}}) \, .
\end{equation}
The positive constant $K_d$ depends only on the dimensionality and the structure of the lattice.
This asymptotic expression does not provide any information on when the asymptotic limit is reached.
Since it has been observed that the Rosenstock approximation describes quite well the high-$\Phi$
regime, it is obvious that a crossover to the Donsker-Varadhan behavior will take place.
The location of this crossover has been studied in detail \cite{Bunde,Gallos1,Gallos2}, and it was shown that only with indirect
methods it is possible to reach the asymptotic limit.

This asymptotic behavior has also been explained via heuristic arguments. The slow relaxation of $\Phi$
at long times is due to an interplay of two different factors. First, mean-field treatments assume
a uniform trap distribution over the entire space. This is not strictly true, though, and for large
enough sizes it is possible to find very extended trap-free regions. A random walker in such a region will
survive for extremely long times compared to walkers in normal regions and will thus determine the
asymptotic behavior. The second factor is due to unusually `compact' random walks, which revisit many
times the same sites, and thus result to a very small value of $S_n$, even at longer times.

Recently, a number of papers were published concerning trapping on a version of the
small-world networks \cite{Jasch2001,Blumen2002,Jasch2002}.
These networks, first proposed as a model by Watts and Strogatz \cite{WS}, are one-dimensional rings
where additional links are inserted between two random sites with a given probability.
It was shown that the results represent a fine interplay between pure order and pure disorder statistics.
Initially, the walkers feel only the presence of the one-dimensional lattice, but at longer
times the behavior of the survival probability follows that of an open tree structure.
The decays of the survival probability were clearly not exponential, and the cumulants description
did not yield accurate coincidence with the numerical results in all of the studied cases.

In this work, we extend the above mentioned random walk problems (mean-squared displacement, coverage, and trapping)
in the case where the underlying structure is a scale-free
network, obeying a power-law in the nodes connectivity distribution. The random walkers are located
on the nodes and can only move along the links of this network. In the case of the trapping model certain nodes
are designated as traps, having a concentration $c$. We present computer simulations results for different
network connectivities and compare them to the known lattice behavior.

\section{The model}

The construction of a scale-free network follows the Molloy-Reed scheme \cite{MR98}: First, we fix the number of nodes $N$
in the system and the $\gamma$ parameter, characteristic of the particular network. By using the
transformation method we select $N$ random numbers from the $k^{-\gamma}$ distribution, so that each node i is
assigned a number of links $k_i$ from the above distribution. The value of $k$ lies in the range from $k_{\rm min}=1$ (lower cutoff) to $k_{\rm max}=N-1$ (no upper cutoff
value is used for $k$).

Initially, no links are established in the system. Each node i extends $k_i$ `hands' towards all other nodes.
We randomly select two such `hands' (that do not belong in the same node) and connect them creating thus a link.
No double links are allowed, so that if two nodes are already connected this link is rejected.
We continue this process until all nodes have reached their pre-assigned connectivity. However, it is possible
that at the last stages of the construction we will reach a dead-end where no further links may be established
according to the above rules. In this case we simply ignore the `hands' that cannot be connected,
since their number is always very small and the structure of the network is not influenced at all.

The largest cluster in the network is identified via the use of a spreading algorithm. We start with a random
node and mark it with a label, say X. We then mark all the nodes connected to this node by X, and proceed iteratively
by labeling their neighbors, etc, until
the whole cluster has been labeled. We then choose another random node that has not been labeled by X, which means
that it belongs to a different cluster. We mark it by Y and again spread this labeling throughout this cluster.
When the entire network has been labeled we can easily identify the largest cluster from its size.

All random walks in this paper take place on the largest cluster of the network via the following algorithm.
We place a random walker on a randomly chosen site i of the largest cluster. This site has a connectivity $k_i$.
At each Monte-Carlo step the walker makes a jump towards a node connected to i
(i.e. nearest neighbor) with probability $1/k_i$. This process gives a Markovian
walk, since each step is independent of all previous steps. Distances on the network are measured according
to the shortest-length path between two nodes, and the displacement $R$ of a walker is calculated
relative to the initial point.

For the trapping problem, we randomly choose a percentage $c$ of the network nodes and designate them as traps.
A random walker is placed on a random non-trap node and performs the procedure described above until it
meets a trap. In this case, it is annihilated and the time $n$ to trapping is recorded.
We repeat the same process for many independent random walkers and different networks, and we construct a histogram
$H(n,c)$ of the number of walks that last exactly $n$ steps. Then, the survival probability is simply given by
\begin{equation}
\Phi(n,c)  = 1 - \frac{1}{M} \sum_{m=1}^{n} H(m,c) \, ,
\end{equation}
where $M$ is the total number of independent random walks sampled.

Typically, 100 different networks with $N=10^6$ nodes were created, and $10^3$-$10^4$ different random origins 
were chosen on each network. Thus, results were averaged over $10^5$-$10^6$ different realizations of the walk.

\section{Results}

\subsection{Mean squared displacement}

\begin{figure}
\begin{center}
\includegraphics{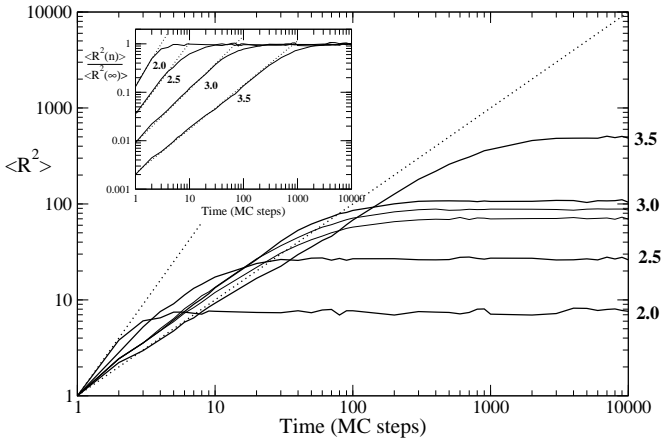}
\end{center}
\caption{ \label{fig_msd} Mean squared displacement as a function of time for networks of $\gamma=2.0$, 2.5,
3.0 and 3.5 (shown on figure). The network size in all cases is $N=10^6$ nodes, and for $\gamma=3.0$
we also present results for networks of size $N=10^4$ and $N=10^5$ (bottom to top). The dotted lines
represent slopes of 1 and 2. Inset: Normalized $\langle R^2(n) \rangle$ curves, so that asymptotically
all curves converge to 1, for different $\gamma$
values (shown on the plot). Dotted lines represent best-fit lines with slopes (left to right)
1.8, 1.5, 1.1, and 0.9.}
\end{figure}

We first study the mean squared displacement $\langle R^2 \rangle$ of a random walker on a scale-free network.
Since the networks that we study are not embedded in a regular Euclidean space, this quantity does not measure
how far in Euclidean space the walker travels, but rather the minimum number of hops needed in order to return to its origin.
The first important feature of Fig.~\ref{fig_msd}, where we present $\langle R^2 \rangle$ as a function
of time for networks of varying connectivity distributions, is the fact that $\langle R^2 \rangle$
equilibrates after a few steps to a constant displacement value. This is a simple manifestation of the very small
diameter of these networks, 
which has been shown to be of the order $\ln(\ln N)$ \cite{Cohen2003}. In practice, this means that one node can
be reached from all other nodes in the network within only a few steps and the maximum possible distance in the
network is very small compared to the network size. Note also that the plateau value increases as we increase $\gamma$, since a network which is
less connected exhibits a larger diameter.
The existence of the plateau is, of course, a finite-size effect. However, the size dependence is not strong,
as can be seen in the figure, where we present results for networks with $\gamma=3.0$ and size $N=10^4$,
$10^5$, and $10^6$. Although we increase the size of the networks by two orders of
magnitude, the value of $\langle R^2 \rangle$ increases from roughly 70 to 110, i.e. the distance $R$
increases almost linearly from 8 to 10. This logarithmic dependence shows that for all practical applications
the plateau will be present. For example, it has been observed \cite{AJB} that the diameter of WWW
(of size $N=8\times 10^8$ and $\gamma=2.45$) is only 18, so that even on such large networks maximum
distances remain small.

In the figure inset we have rescaled the $\langle R^2(n) \rangle$ data so that all curves are
normalized to an asymptotic value of 1. It is shown that upon varying the value of $\gamma$,
diffusion on scale-free networks may range from superlinear to sublinear diffusion.
For networks of low $\gamma$, diffusion is greatly enhanced.
Thus, for $\gamma=2.0$ the walkers move away from
the origin rapidly and the slope of $\langle R^2 \rangle$ reaches a value of about 1.8.
After only a few steps, though, the value of $\langle R^2 \rangle$ saturates, due to the phenomenon described
above. As we increase the value of $\gamma$ the slope of the curves decreases. Diffusion at
early steps remains super-linear, until we reach a value of $\gamma$ around 3.0 where the slope
becomes roughly equal to 1. This linear diffusion turns slowly into sub-linear as we further increase
$\gamma$ and for $\gamma=3.5$ the slope is equal to 0.9.

\begin{figure}
\begin{center}
\includegraphics{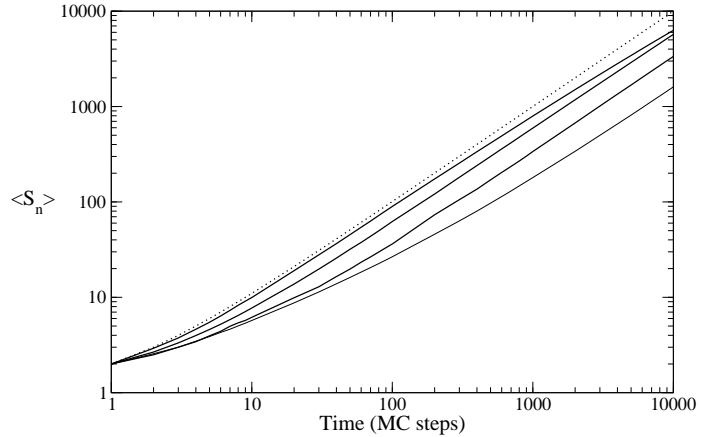}
\end{center}
\caption{ \label{fig_sn} Number of distinct sites visited $\langle S_n \rangle$ after $n$ steps on scale-free networks
with (solid lines, top to bottom) $\gamma$=2.0, 2.5, 3.0 and 3.5. The dashed line is the infinite-dimensional case
of $\langle S_n\rangle=n+1$. The network size was $N=10^6$.}
\end{figure}

\subsection{Network coverage}

The coverage of the network by random walks is found to be a very efficient process.
Numerical results of $\langle S_n\rangle$ on scale-free networks are presented in Fig.~\ref{fig_sn}.
The number of sites visited increases initially with a slow rate, but after a crossover
value the increase is almost linear. This asymptotic linearity is observed in all $\gamma$ values,
while the crossover point shifts towards longer times with increasing $\gamma$.
The early time slope means that the walkers initially spend some time exploring the neighborhood they were
created in and visit the same sites. After the first few steps (the exact number depends on the connectivity
of the network) they escape their initial territory and diffuse around the entire network. Thus,
it is possible to continuously visit new sites, which results in the linear increase of $\langle S_n\rangle$.
In reference \cite{Almaas}, $\langle S_n\rangle$ on small world networks was found to scale at early times 
with $\sqrt{n}$ and asymptotically with $n$.
In the case of scale-free networks, the early-time behavior is not consistent with a $\sqrt{n}$ law, which
would be an indication of one-dimensional behavior. For each $\gamma$ value the local environment is different,
and this is exhibited in the different evolution of $\langle S_n\rangle$ for low $n$ values.
The crossover, also, is located at much earlier times
(of the order of tens of steps) as compared to thousands of steps which is the case reported in \cite{Almaas}
for small world networks.

The size of the network used ($N=10^6$) was two orders of magnitude larger than the number of steps performed,
in order to avoid finite size effects. Despite of this precaution, the curve of $\gamma=2$ seems to deviate from linearity at longer
times. This phenomenon means that revisitation already starts to exhibit itself for the finite
network we study.

The linear growth of $\langle S_n\rangle$ is similar to the behavior exhibited in dendrimer structures, modeled by Cayley trees.
These are open structures, with every node having a fixed number $k$ of connected nodes which are always directed
away from the central core. It was also shown in that case \cite{Katsoulis} that $\langle S_n \rangle$ had a linear increase after a short
early-time sublinear regime, due to the same reasons as here.

Consider, now, a regular lattice that can be embedded in a finite $d$-dimensional space. In this case,
it is well-known that unconstrained diffusion causes the random walker to spread in the available space, increasing
both $\langle R^2 \rangle$ and $\langle S_n \rangle$ with time. Thus, the walkers tend to increase
their distance from the origin and cover new territory. If for some reason we restrict diffusion of
the walkers within a finite distance $R_c$ from the origin, so that $\langle R^2 \rangle$ saturates,
then the area covered will soon saturate, too, to a value of order $R_c^d$. Diffusion on scale-free
networks, though, is different. Although the walker
is always close to the origin and restricted within a distance equal to the network diameter,
new territory is continuously sampled. This peculiar behavior can be attributed to
the existence of the hubs. If we consider an extreme case of a hub, that of a star node where all the
nodes are connected only to the hub, then the displacement will be at most 2 steps away, but due to the large number
of nodes in the system the walker will be redirected to a non-visited site with a revisitation probability
$n/N$, which for large enough systems and early to moderate times is close to 0.
The particular case of L\'{e}vy flights \cite{Levy} (which usually results in enhanced diffusion with $a>1$ in Eq.~\ref{EQmsd})
can be considered similar to the process we study in this paper. In a L\'{e}vy flight the length of a jump follows a power law
dependence. In practice, a walker samples an area for a certain amount of time before performing a long-range jump.
This jump allows then the walker to sample a new space. Although the areas of space visited can be regarded as
the hubs of the present problem, the main difference is the displacement of the walk. In the case of L\'{e}vy flights
the `hubs' of the system are distant in space among each other, while in our case the hubs are very close
to each other, and can even be directly connected. As we have already seen, thus,
the mean squared displacement on scale-free networks (even of large size) is restricted to small distances, whereas $\langle R^2 \rangle$
increases monotonically in the case of L\'{e}vy flights.

\subsection{Trapping}

\begin{figure}
\begin{center}
\includegraphics{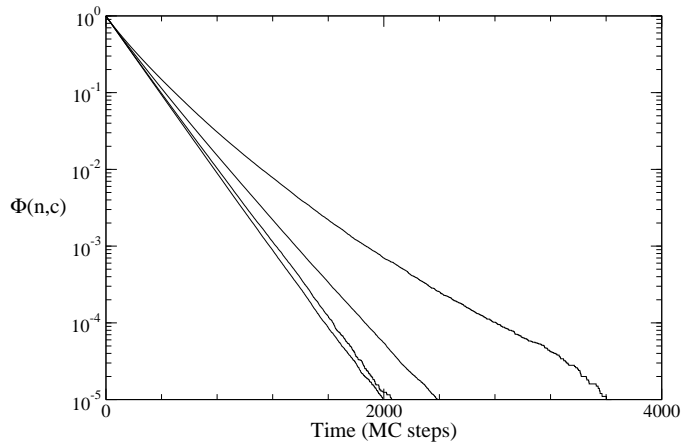}
\end{center}
\caption{ \label{N_dependence} Survival probability as a function of time, for a network of 
$\gamma=2.5$ and trap concentration $c=0.01$. From right to left, the number of nodes
in the network is $N=10^3$, $10^4$, $10^5$, and $10^6$.}
\end{figure}

For the trapping problem, we first examine the dependence of the survival probability $\Phi$
on the system size $N$. As it can be seen in Fig.~\ref{N_dependence}
for $\gamma=2.5$, larger networks yield a significantly lower survival probability. This is due to
the higher probability of finding a node with very high connectivity, which is linked
directly with the largest part of the network. Due to the power-law dependence the appearance of these nodes
increases as we increase the network size. However, we can see that the $\Phi$-curves for the larger
networks ($N\geq 10^5$) practically coincide. Moreover, this $N$-dependence is much
weaker for networks with higher $\gamma$ values. As we can see in the plot, the large-size network behavior
in this case is very close to a simple exponential decay, while smaller networks deviate from this behavior.

\begin{figure}
\begin{center}
\includegraphics{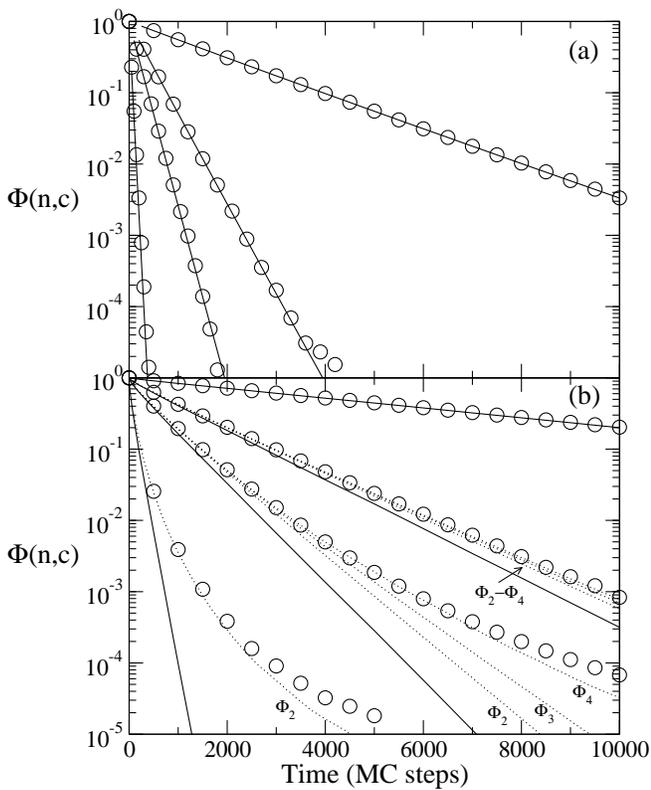}
\end{center}
\caption{ \label{c_dependence} Survival probability of random walkers as a function of time, for a network of size $N=10^6$ and
(a) $\gamma=2.5$, (b) $\gamma=3.5$. Symbols represent direct trapping simulations.
Solid lines represent the Rosenstock approximation based on the $\langle S_n \rangle$  data of figure \protect\ref{fig_sn}.
Dashed lines are the results of the cumulant approximation, with the truncation order $J$ indicated on the plot.
From left to right, the trap concentrations are $c=$ 0.05, 0.01, 0.005, and 0.001.}
\end{figure}

In Fig.~\ref{c_dependence}(a) we present the survival probability on the
largest cluster for different trap concentrations as a function of time, for networks with
$\gamma=2.5$. For a relatively high trap concentration, e.g. $c=0.05$, we
can see that $\Phi$ falls very rapidly and during the first 100 steps only a small percentage of the walkers has
survived. The decay retains for the largest part an exponential character.
In order to test the validity of the Rosenstock approximation for scale-free networks, we
used the numerical data for $\langle S_n\rangle$ presented in Fig.~\ref{fig_sn} and computed the survival probability $\Phi$
using Eq.~(\ref{Rosen}). The results in Fig.~\ref{c_dependence} show that there is 
almost complete coincidence between this approximation and the simulation data.
As we have mentioned above, the Rosenstock approximation is valid when mean-field features are present, and
fluctuations in the area covered are not important.  Thus, a high trap concentration implies that
there will be no large trap-free regions, since a walker can easily escape any part of the system.
However, in the case of $\gamma=2.5$ the same argument is true as we gradually move towards lower concentrations.
The survival probability retains the simple exponential character as we decrease $c$, even for the lowest trap concentrations used.
The Rosenstock approximation, Eq.~(\ref{Rosen}), predicts this simple exponential decay only in the time range where
$\langle S_n \rangle \sim n$. As we have seen, though, in Fig.~\ref{fig_sn} there is a crossover in the behavior of
$\langle S_n \rangle$ with time, which should modify this behavior. However, this crossover takes place at early times
and is not apparent in the linear time scale used for the survival probability. The Rosenstock approximation, thus,
based on the results of Fig.~\ref{fig_sn}, predicts a simple exponential decay for $\Phi$ on scale-free networks.

Fig.~\ref{c_dependence}(a) validates, thus, the assumption that the Rosenstock approximation is true
in the case of scale-free networks with $\gamma=2.5$.
The decay of the survival probability, though, is greatly influenced by $\gamma$. In Fig.~\ref{c_dependence}(b),
the results for $\gamma=3.5$ and large trap concentrations $c$ clearly demonstrate a deviation from a simple exponential behavior
and the failure of the Rosenstock approximation. Only in the case of low $c$, such as $c=10^{-3}$, this approximation
is satisfactory and describes reasonably well the exponential decay of the simulation data. 
Thus, for $\gamma=3.5$ we also employed the cumulant approximation of Eq.~(\ref{EQ_cum_surv}). The higher-order
moments of the $S_n$ distribution were calculated numerically, via the same simulations that yielded the
first moment $\langle S_n\rangle$ of Fig.~\ref{fig_sn}. It is evident
that the description of the data improves significantly. The second-order truncation (i.e. including the
standard deviation of the $S_n$ distribution) follows quite closely the simulation data for $c=0.05$ and $c=0.005$
over more than three decades on the vertical axis.
In the case of $c=0.01$ we need to include higher moments in order to achieve the same level of accuracy,
since $\Phi_2$ captures only part of the behavior. The fourth-order truncation $\Phi_4$ seems to be quite
succesful over almost four decades, and describes a significant non-exponential part of the curve quite well.

\begin{figure}
\begin{center}
\includegraphics{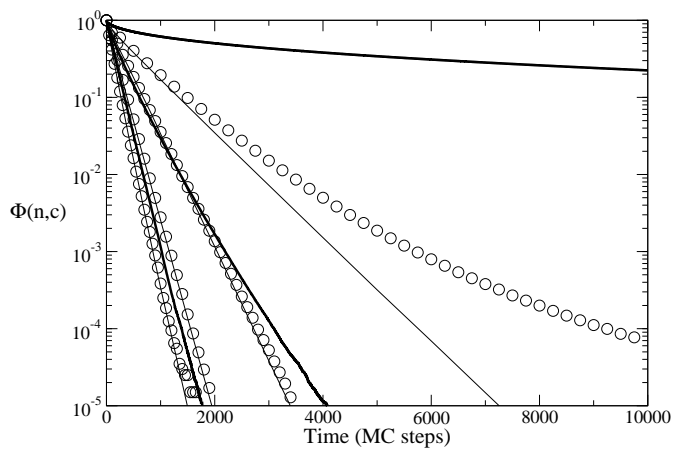}
\end{center}	
\caption{ \label{g_dependence} Survival probability as a function of time, for networks
of size $N=10^6$ and trap concentration $c=0.01$. From left to right, the network connectivity
is (symbols) $\gamma=$ 2.0, 2.5, 3.0 and 3.5. The corresponding Rosenstock approximation is represented with thin lines.
We also present the survival probability for
regular networks (thick lines) in (top to bottom) $d=$ 1, 2 and 3 dimensions.}
\end{figure}

In Fig.~\ref{g_dependence} we present the survival probability ($c=0.01$) in different scale-free networks and in
regular lattices. We can see that as $\gamma$ increases the survival probability becomes higher. Since 
the number of connections between the nodes decreases with $\gamma$ and we have seen that the average value
of the number of sites visited also decreases, the walkers will spend more time in smaller network
regions. This has a dramatic influence on $\Phi$ and as we can see in the figure the difference in the
survival probability between networks of $\gamma=2$ and $\gamma=3.5$ can be two orders of magnitude,
even only after a few hundred steps. The shape of the curves is also different, since the
exponential character of the lowest $\gamma$ values is no longer retained for $\gamma>3$. This change
in the decay, along with the much slower relaxation is a manifestation of the network structure,
which for $\gamma>3$ corresponds to a loosely connected network where the number of nodes with extremely high
connectivity has diminished.

Inspection of Fig.~\ref{g_dependence} and similar simulations for different concentrations on networks
with $\gamma=3$ suggest that in the range $2<\gamma<3$ the Rosenstock approximation provides a reliable description
in the time regime studied in this work. On the contrary, when $\gamma>3$ this approximation is not valid and one needs to
resort to the use of higher moments in the cumulant expression.

Concerning the comparison with regular lattices, it is obvious that trapping in the most connected networks
($\gamma=$2-3) behaves in a similar manner as in 3-dimensional lattices (simple exponential decay), and for
$\gamma\leq 2.5$ decays in a similar rate, too. The case of a two-dimensional lattice represents the borderline
dimension for recurrent random walks in lattices, and the relaxation of $\Phi$ is not exponential,
while for $d=1$ the survival probability is considerably higher, since the walkers are confined between
two trapping sites and perform a random walk in this region. Similarly to the $d\leq 2$ cases, the survival
probability relaxation in networks with $\gamma>3$ is not exponential and, in general, cannot be described
by the Rosenstock approximation.

Scale-free networks have been considered heuristically to behave as infinite-dimensional lattices. This assumption
($d\to \infty$), however, implies that both the Rosenstock approximation (Eq.~\ref{Rosen})
and the Donsker-Varadhan result (Eq.~\ref{dv}) would
yield a single exponential decay $\Phi(n) \sim \exp(-n)$ with the number of steps $n$. As we have seen, though,
this result can be verified in the presented time scale by the simulations for networks in the range $2<\gamma<3$, but not for $\gamma>3$.
The reason is that in $d\to \infty$ the probability
for a walker to revisit a site is vanishingly small, since at every step the walker has an infinite number
of possible sites to jump to. Thus, the revisitation probability tends to zero and the number of sites visited is
equal to the number of steps performed ($\langle S_n \rangle\sim n$). Eq.~(\ref{Rosen}) then predicts the same behavior
as (\ref{dv}), i.e. $\Phi(n)\sim \exp(-n)$. For scale-free networks the situation
is somewhat different, though.
Although there are a few highly connected nodes in the system (hubs), from where a walker can be directed to
previously unsampled areas of the network, the largest percentage of the nodes has a very small number of links, e.g. $k=1$ or $k=2$.
A walker that reaches such a node will return at the next step to its former position. The character of a scale-free
network as a substrate for random walks, thus, cannot be described as purely infinite-dimensional.
The dimensionality can be considered as a local property which is modified according to the area of the
network where the walker lies in. Depending on the value
of $\gamma$, the area sampled depends on how connected a system can be and how easy it is for a random walker to
visit new nodes.
For sparse networks, for example, the revisitation probability increases (together with the network diameter)
and leads to larger deviations of the above law.

\section{Summary}

In this work we presented numerical results on $\langle R^2 \rangle$, $\langle S_n \rangle$,
and trapping in scale-free networks, which are well-studied processes in many other systems. 

Mean squared displacement was found to range from superlinear diffusion to sublinear diffusion
as we varied $\gamma$, while the network coverage increases almost linearly with time for all $\gamma$
values examined. The Rosenstock approximation is adequate for predicting the
survival probability in the range $2<\gamma<3$, but for higher $\gamma$ it cannot account
for the non-exponential character of the survival probability decay with time. In this
case, we found that the cumulant expansion can fit quite accurately the observed behavior.

The mean-field character (exhibited by the validity of the Rosenstock approximation) for $\gamma<3$
can be also expected in the case of these networks, since the heuristic
arguments supporting the Donsker-Varadhan expression (Eq.~\ref{dv}) do not apply here. Although a
walk can still be compact, large trap-free regions do not exist on such a network. The main reason
is the small average path length between any two nodes of the structure. For any trap distribution, there are not any
network areas where a walker can spend a lot of time without meeting a trap, since the connectivity
of the network allows it to easily escape to a different neighborhood, where traps may exist in a larger
local concentration. When $\gamma>3$, though, the importance of the hubs in a network diminishes,
and the behavior resembles more that of regular low-dimension lattices, with prominent non-exponential
decays even at early times.

\end{document}